\documentclass[letterpaper,english,reprint,aps,prl,superscriptaddress,showpacs]{revtex4-1}
\usepackage[latin9]{inputenc}
\usepackage{amsmath}
\usepackage{amssymb}
\usepackage{graphicx}

\makeatletter

\pdfpageheight\paperheight
\pdfpagewidth\paperwidth

\newcommand{\lyxdot}{.}

\usepackage{babel}
\usepackage[
colorlinks,
linkcolor=blue,
anchorcolor=black,
urlcolor=blue,
citecolor=blue
]{hyperref}

\makeatother

\usepackage{babel}
\begin{document}
\title{Epidemic spreading and herd immunity in a driven non-equilibrium system of strongly-interacting
atoms}
\author{Dong-Sheng Ding}
\email{dds@ustc.edu.cn}

\affiliation{Key Laboratory of Quantum Information, University of Science and Technology
of China, Hefei, Anhui 230026, China.}
\affiliation{Synergetic Innovation Center of Quantum Information and Quantum Physics,
University of Science and Technology of China, Hefei, Anhui 230026,
China.}
\author{Zong-Kai Liu}
\affiliation{Key Laboratory of Quantum Information, University of Science and Technology
of China, Hefei, Anhui 230026, China.}
\affiliation{Synergetic Innovation Center of Quantum Information and Quantum Physics,
University of Science and Technology of China, Hefei, Anhui 230026,
China.}
\author{Hannes Busche}
\affiliation{Department of Physics, Chemistry and Pharmacy, Physics@SDU, University of Southern Denmark, 5230 Odense M, Denmark }

\author{Bao-Sen Shi}
\email{drshi@ustc.edu.cn}

\affiliation{Key Laboratory of Quantum Information, University of Science and Technology of China, Hefei, Anhui 230026, China.}
\affiliation{Synergetic Innovation Center of Quantum Information and Quantum Physics, University of Science and Technology of China, Hefei, Anhui 230026,
China.}
\author{Guang-Can Guo}
\affiliation{Key Laboratory of Quantum Information, University of Science and Technology
of China, Hefei, Anhui 230026, China.}
\affiliation{Synergetic Innovation Center of Quantum Information and Quantum Physics, University of Science and Technology of China, Hefei, Anhui 230026,
China.}
\author{Charles S. Adams}
\email{c.s.adams@durham.ac.uk}

\affiliation{Department of Physics, Joint Quantum Centre (JQC) Durham-Newcastle,
Durham University, South Road, Durham DH1 3LE, United Kingdom.}
\author{Franco Nori}
\email{fnori@riken.jp}

\affiliation{Theoretical Quantum Physics Laboratory, RIKEN Cluster for Pioneering Research, Wako-shi, Saitama 351-0198, Japan.}
\affiliation{RIKEN Center for Quantum Computing (RQC), Wako-shi, Saitama 351-0198, Japan.}
\affiliation{Physics Department, The University of Michigan, Ann Arbor, MI 48109-1040,USA.}
\date{\today}

\maketitle
\textbf{It is increasingly important to understand the spatial dynamics of epidemics. While there are
numerous mathematical models of epidemics, there is a scarcity of physical
systems with sufficiently well-controlled parameters to allow
quantitative model testing.  It is also challenging to replicate the
macro non-equilibrium effects of complex models in microscopic systems.
In this work, we demonstrate experimentally a physics analog of epidemic spreading
using optically-driven non-equilibrium phase transitions in strongly
interacting Rydberg atoms. Using multiple laser beams we can impose
any desired spatial structure. We observe spatially localized
phase transitions and their interplay in different parts of the sample.
These phase transitions simulate the outbreak of an infectious disease in
multiple locations, as well as the dynamics towards ``herd immunity" and ``endemic state" in different regimes. The reported results indicate that Rydberg systems
are versatile enough to model complex spatial-temporal dynamics.}

Self-organization and non-equilibrium dynamics of complex systems emerge not only in physics, but also in other fields such
as earth science, biology, and economics \citep{haken2006information}.
In these cases, many interacting degrees of freedom are repelling,
such as sand and rice grains \citep{per1987self}, vortices \citep{field1995superconducting,olson1997superconducting},
etc. and they can be temporally stuck in metastable states, due to local
traps or due to static friction. When slowly driven towards marginal
stability, these extended systems exhibit avalanches at various length scales \citep{per1987self, field1995superconducting,olson1997superconducting}.
Rydberg atoms allow to model spatial dynamics with emerging complexity \citep{schauss2012observation,labuhn2016tunable,bernien2017probing,de2019observation,bluvstein2021controlling}.
These atoms combine the precision of atomic physics with strong interactions
with neighboring atoms, more common in condensed matter systems \citep{eisert2015quantum,browaeys2020many}.
Rydberg atoms are not only ideal for applications in quantum technology
and computing \citep{saffman2010quantum,Adams_2019,morgado2020quantum},
but also as a potential model system for simulations \citep{bloch2012quantum,georgescu2014quantum,morgado2020quantum}.
Precise control of the excitation probability using narrowband lasers
\citep{gallagher2005rydberg,firstenberg2016nonlinear} makes Rydberg
atoms an excellent candidate to study non-equilibrium physics, as
instabilities of equilibrium states are ubiquitous processes occurring
in a variety of driven Rydberg systems. For example, Rydberg atoms
in different driven configurations exhibit many fascinating characteristics
of complex systems, including aggregate formation \citep{urvoy2015strongly},
non-equilibrium phase transitions \citep{carr2013nonequilibrium,malossi2014full},
critical points \citep{marcuzzi2014universal}, self-organized
criticality \citep{helmrich2020signatures,ding2020phase}, epidemic
growth \citep{wintermantel2021epidemic}, and hydrodynamics
\citep{klocke2021hydrodynamic}.

There is considerable interest in simulating the outbreak of infectious
diseases. The accuracy of epidemic models has been considerably improved
over time \citep{eubank2004modelling,longini2005containing}. It is desirable to study and replicate the macro non-equilibrium effects of complex dynamics in well-controlled microscopic atomic systems. Moreover, many non-equilibrium phenomena cannot be broken down to just two global stable states; and numerous real systems are spatially inhomogeneous, e.g. the outbreak of an infectious disease can occur in multiple locations \citep{pastor2015epidemic}.
The added interplay between multiple spatial domains results in additional complexity that manifests as both local and non-local dynamics. Consequently, the study of phase transitions in engineered spatially inhomogeneous systems is an important tool to improve our understanding of non-equilibrium dynamics in extended systems.

\begin{figure*}[t]
\includegraphics[width=2.0\columnwidth]{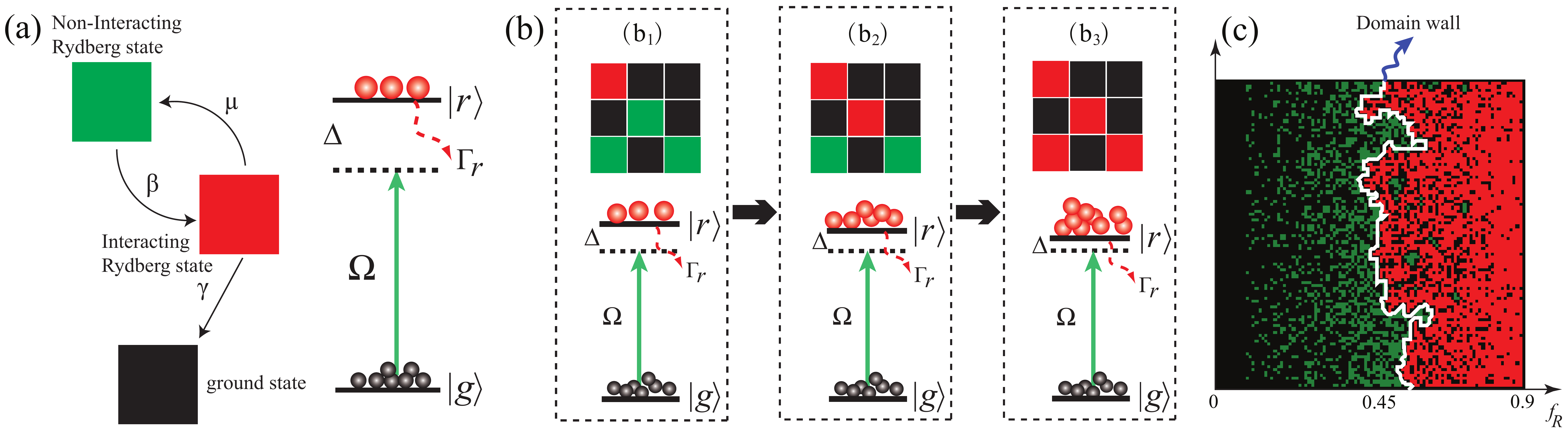}\caption{\textbf{Epidemic model simulation with Rydberg avalanche excitations.} (a) The model consists of three states: the interacting Rydberg state (red), the non-interacting Rydberg state (green) and the ground state (black). The right diagram shows the energy levels of atoms under a laser drive. Here, $\Delta$ is defined as the effective detuning between the Rydberg state $\lvert r\rangle$ and the laser, and $\Gamma_{r}$ is the decay rate of $\lvert r\rangle$. (b) The algorithm of our 2D cellular automaton. The rule of updating cells in the susceptible-infected-recovered model is shown in
(b$_1$--b$_3$): the cells are updated in the next step according to its current state and the state of its eight nearest neighbors. The susceptible infected-susceptible model satisfies the condition $\beta\gg\gamma$, $\gamma\rightarrow0,\mu=0.01$; while the susceptible-infected-recovered process obeys $\beta/\gamma>1$ and $\mu\rightarrow0$. The interacting Rydberg atoms spread out in the entire interacting region because the population-dependent energy shift $\Delta_{\rm shift}$ induces a nonlinear `facilitated' excitation, in which the effective detuning $\Delta$ between the laser and the Rydberg energy level of the excited state $\lvert r\rangle$ is shifted. (c) A 2D snapshot of interacting Rydberg atoms (red) spreading in a $N=100\times100$ inhomogeneous Rydberg density system (with fractions $f_{R}=0$--$0.9$) after $i=200$ iterations. There is an isolated boundary of contagion (marked in white), a domain wall separating the inhomogeneous distributions of Rydberg density.}
\label{avalanche results}
\end{figure*}

Here we model epidemic spreading dynamics using a laser-driven thermal
Rydberg gas. We have observed the analog of typical transmission profiles
for susceptible infected-susceptible (SIS) and susceptible-infected-recovered
(SIR) processes. The intensity pattern of the laser beams creates a spatially
inhomogeneous Rydberg excitation rate, and thus spatial domains with
different Rydberg atom densities. Within each domain, the system can
be found in one of two phases. Below a critical threshold density,
interactions between Rydberg atoms are negligible; We refer to this
as the \textit{`non-interacting'} NI phase. Above the threshold,
interactions between Rydberg atoms induce a shift and broadening of
the Rydberg lines that facilitate Rydberg excitation in adjacent regions;
This triggers a localized excitation avalanche and transitions to
a \textit{`strongly-interacting'} I phase. We observe separate
phase jumps and hysteresis loops associated with different spatial
domains within our medium. The multiple hysteresis loops combine to
produce multistability in the optical transmission. The experimental atomic system shows an initial exponential spread of the (microscopic) ``epidemic", until the system reaches two cases: one is ``herd immunity", which stops the spread in the SIR process; another is ``endemic state", which is a stationary state in the SIS case.

In order to gain insight into the spatial dynamics of Rydberg non-equilibrium
phase transitions and, in particular, the spreading of the interacting phase,
we build an avalanche model via the SIS and SIR models introduced
in Refs.~\citep{pastor2015epidemic}. The elements of the effective model
are schematically shown in Fig.~\ref{avalanche results}(a); where $\beta$
is the spreading rate from the non-interacting Rydberg state to the
interacting Rydberg state, $\mu$ is the rate of the reverse process, and $\gamma$
% $\Gamma_{\rm r}$
is the decay rate of the interacting Rydberg state to the ground state.
The spreading process is simulated by a cellular automaton with the
rules shown in Fig.~\ref{avalanche results}(b). The model uses
$N=m\times m$ 2D cells. Each cell updates according to its current
state and the state of its eight nearest neighbors.
First, the cells are randomly filled with fractions $f_{R}$ of Rydberg atoms. A cell is in the NI phase if $f_{R}<f_{R,c}$ [green cells
in Fig.~\ref{avalanche results}(b)], in the I phase if $f_{R}>f_{R,c}$
(red), or in a depleted phase (black) without Rydberg atoms in the
cell. These situations correspond to a susceptible person, an infected
person, or an immune site, in the original epidemic model \citep{pastor2015epidemic}.

The procedure schematically shown in Fig.~\ref{avalanche results}(b) describes the following steps: if the fraction $f_R$ of Rydberg atoms in an arbitrary cell [red cells in figure~\ref{avalanche results}(b$_1$)] exceeds the critical fraction $f_{R,c}$, and is thus in the interacting phase, it increases the excitation probability of neighboring Rydberg atoms [green cells in Fig.~\ref{avalanche results}(b$_1$)] $f_{R}$ to be above $f_{R,c}$, such that the neighboring cells transition to the interacting phase as well in the next iteration [red cells in Fig.~\ref{avalanche results}(b$_2$)].
This occurs due to level shifts $\Delta_{\rm shift}$
by strong interactions that result in a `facilitated' excitation process, which triggers an avalanche process throughout the interaction region, corresponding to the epidemic spreading to adjacent sites; see the energy diagrams in Fig.~\ref{avalanche results}(b$_1$--b$_3$). We consider the condition $\beta\gg\gamma,\mu$ for the following SIS simulation, which corresponds to fast scanning either the probe intensity or $\Delta_{c}$, and the system changes from the NI to the I phase. Otherwise, the system would oscillate
near the critical point, and the oscillations between the phases display a bimodal distribution of transmission levels as demonstrated in Ref.~\citep{ding2020phase}, see also \citep{supplement}. The boundary condition in our model is such that interacting Rydberg atoms at the edges disappear as they would move out of the excitation volume defined by the laser beams.

\begin{figure}
\includegraphics[width=1\columnwidth]{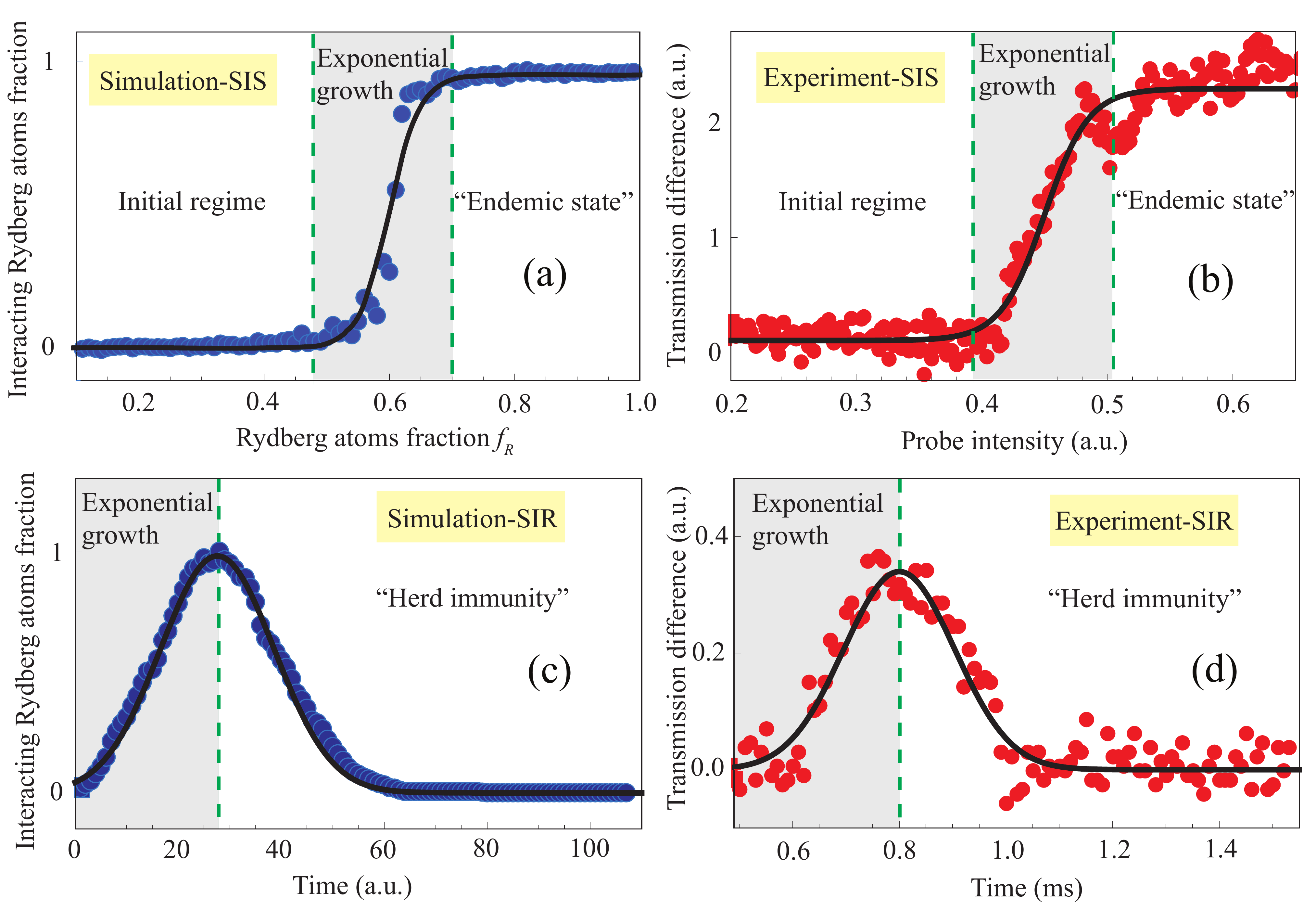}\caption{\textbf{Epidemic nonlinear spreading}
(a) SIS model by simulating avalanche behavior of interacting Rydberg atoms via a cellular automaton in an $N=100\times100$ cells system. (b) Measured phase transition versus probe intensity. The solid lines in (a) and (b) are fit by the hyperbolic tangent function $A+B\tanh[(x-C)/\omega]$, with parameters $A=B=0.47$, $C=0.6$ and $\omega=0.05$ for (a), and $A=0.12\text{, }B=0.11$, $C=0.45$ and $\omega=0.035$ for (b). (c) SIR model showing a peak versus time. (d) The measured transmission in time without scanning laser detuning. The solid lines in (c) and (d) are fit by the Gaussian function $Be^{[-\omega(t-C)^{2}]}$ with parameters $B=0.98$, $C=27.6$ and $\omega=0.0041$ for (c), and $B=0.034$, $C=0.8$ and $\omega=46$ for (b). The dynamics is divided in three regimes: initial, exponential growth, and ``herd immunity" or ``endemic state". The gray areas in (a-d) are the exponential growth regions (or nonlinear outbreak) in the epidemic process.}
\label{phase transition}
\end{figure}

We have simulated the dynamics of `facilitated' Rydberg excitations
according to the cellular automaton, see Fig.~\ref{avalanche results}(c).
After $i=200$ iterations in a 2D inhomogeneous Rydberg density system
from fractions $f_{R}=0$ to $f_{R}=0.9$, respectively, with $m=100$, there is an isolated boundary edge [shown in white in Fig.~\ref{avalanche results}(c)]  or ``domain wall" that splits up the NI and I phases.
On the contrary, there are no obvious domain walls if using a uniform
atomic density. In the latter case, we record the Rydberg atoms density
in the I phase after $200$ iterations, and the simulated result
is shown in Fig.~\ref{phase transition}(a). The interacting Rydberg fraction versus the Rydberg density $f_R$ shows a critical point near a Rydberg fraction $f_{R,c}=0.55$--$0.65$. In the experiment, we measure the transmission of the probe beam against the Rydberg atoms fraction by increasing the probe intensity in a two-photon electromagnetically induced transparency (EIT) scheme, see more details on the experimental
configuration in \citep{supplement}. There is a sudden jump in the probe transmission spectrum given in Fig.~\ref{phase transition}(b), which corresponds to the system transition from the NI to the I phase within the exponential growth regime. The threshold effect in the spreading of the interacting Rydberg atoms is consistent with the nonlinear spreading characteristics in the SIS model, as predicted in Fig.~\ref{phase transition}(a). In the case, the system is scanned in a fast rate (see more details in \citep{supplement}), the NI and I phases tends to dynamical equilibrium, which predicts a stationary ``endemic state", a main character of SIS model. When we consider the SIR process, the parameters $\beta$ and $\mu$ satisfy $\beta/\gamma>1$ and $\mu\rightarrow0$, corresponding to the case of measuring the Rydberg excitations without scanning the laser detuning in the experiment. We have simulated the results in Fig.~\ref{phase transition}(c) with $\beta=0.95$ and $\gamma=0.2$. The interacting Rydberg atoms fraction first increases due to the contagion effect in the exponential regime, and then suddenly decreases because the system reaches \textit{``herd immunity"} via the ``infection recovery channel" in the SIR process. We have also measured the probe transmission spike near the critical point in the time domain, as shown in Fig.~\ref{phase transition}(d).
The experimental Rydberg system shows the exponential spread of the (microscopic) epidemic until it reaches ``herd immunity".

\begin{figure}[b]
\includegraphics[width=1\columnwidth]{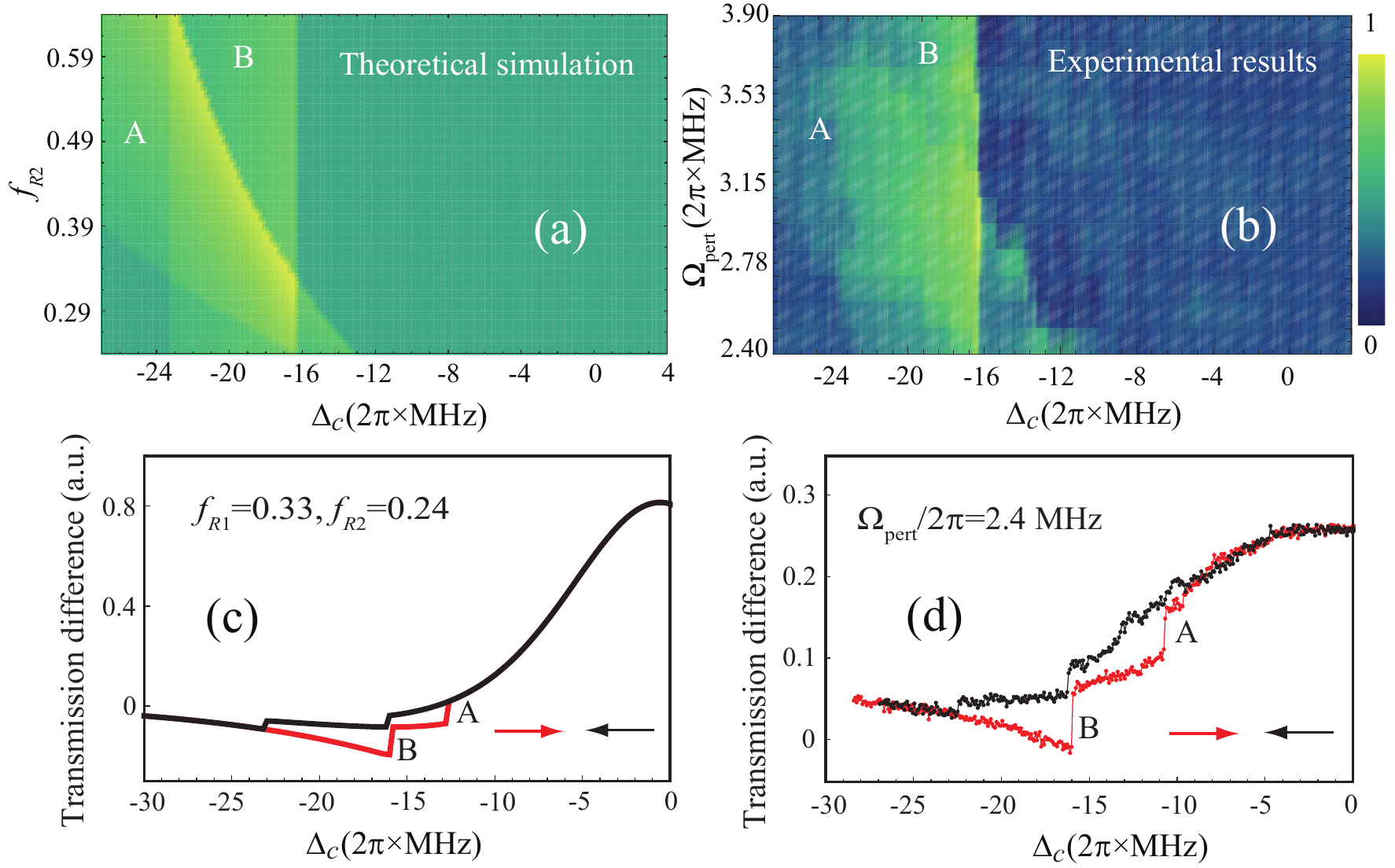}\caption{\textbf{Multi-domain bistability.} Comparison of the (a) theoretically, and (b) experimentally observed probe transmission. The probe transmission versus the coupling detuning $\Delta_{c}$
are shown in (c) theory and (d) experiments.
The two Rydberg fractions $f_{R1}$ and $f_{R2}$
are fixed in (c) and the perturbing field Rabi frequency $\Omega_{\textrm{pert}}$ is fixed in (d). The areas indicated by A and B show their individual bistability regions, which add up to multi-stability where they overlap. The hysteresis loops of multistability are shown in (c) theory and (d) experiment. Positive ($+$) and negative ($-$) scan directions are shown in red and black, respectively. \label{multibistability-tuning}}
\end{figure}

\begin{figure*}[t]
\includegraphics[width=2\columnwidth]{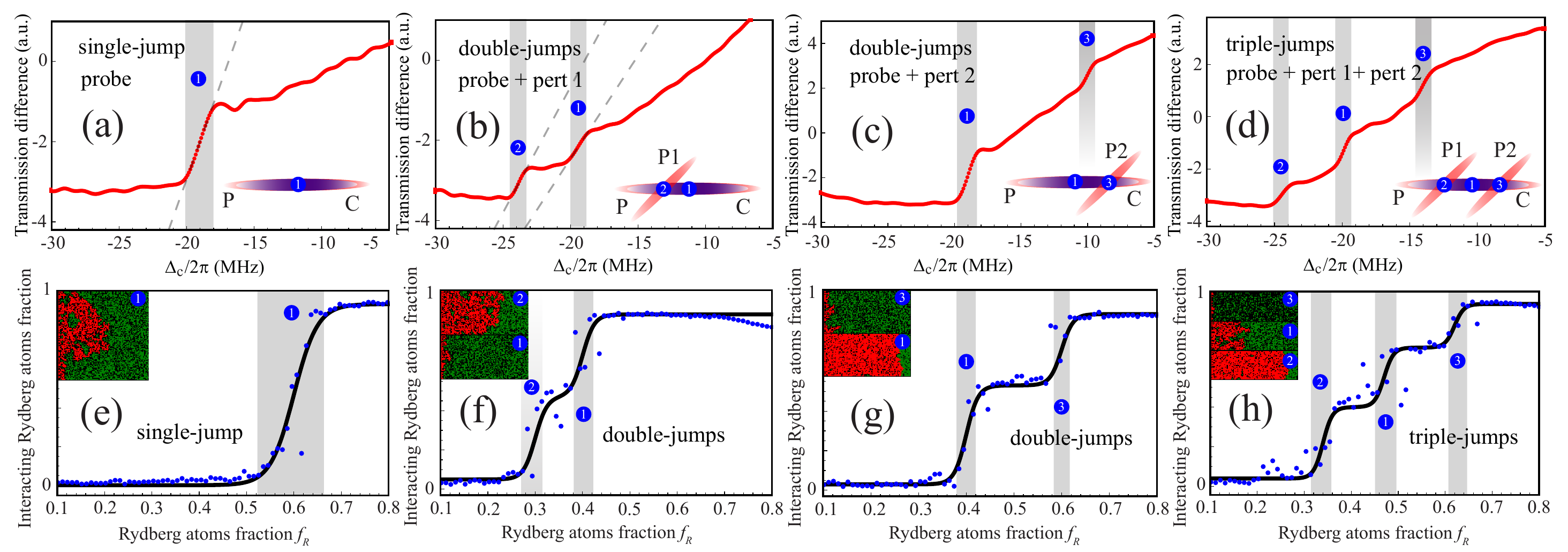}\caption{\textbf{Multiple phase jumps.}
Panels (a)-(d) show, respectively, spectra with single-, double- and triple-jumps
in the transmission. The different spectra are obtained
by adding the additional perturbing fields 1 and 2 to create two and
three domains. In these cases, the angles between the probe and the
two perturbing fields are 7$^{\circ}$ and 14$^{\circ}$, respectively. The dashed lines given in (a) and (b) are the slopes of the jumps. Here, $\Omega_{\textrm{pert,1}}=2\pi\times3.7$~MHz
and $\Omega_{\textrm{pert,2}}=2\pi\times2.2$~MHz. Panels (e)-(h)
show our simulations using the SIS cellular automaton \citep{drossel1992self,ding2020phase}
with one, two, and three spatial domains, respectively. The solid lines
in these simulations are fit by the function $A+\sum B_{i}\tanh((x-C_{i})/\omega_{i})$
including a series of tanh steps with different position $C_{i}$
and width $\omega_{i}$. We initialize the domain with atomic fractions
$f_{R1}=f_{R}$ in (e) and $f_{R1}=f_{R}+0.2$, $f_{R2}=f_{R}+0.3$
in (f), $f_{R1}=f_{R}+0.2$ and $f_{R3}=f_{R}$ in (g) and $f_{R2}=f_{R}+0.3$,
$f_{R1}=f_{R}+0.15$, $f_{R3}=f_{R}$ in (h). The insets show the
spread of interacting atoms (red) from left to right following 100
iterations. }

\label{multi stability}
\end{figure*}

The nonlinear spreading of the I phase is strongly dependent
on the initial Rydberg density $f_{R}$($t$=0). This reveals that one
uniform spatial domain leads to one jump, and multiple spatial domains
lead to multiple jumps. This corresponds to the scenario of an outbreak
of an infectious disease in different locations.
In the experiment, a spatial inhomogeneity is introduced by subjecting
part of the atomic vapor to localized perturbing beams. The additional
perturbing fields have the same frequency as the optical probe field,
but intersect the latter at an angle, creating regions with different
intensities, as explained in \citep{supplement}. Note that, a domain
boundary can also arise from the attenuation of the probe light along
with the medium \citep{carr2013nonequilibrium}. For any spatially
varying intensity distribution, the distinct phases are characterized
by a spatial boundary between the states of high and low population.
The occurrence of optical multistability and non-equilibrium phase
transitions in a multi-domain system can be simulated using an EIT
master equation with a mean-field model, see \citep{supplement}.
The multiple discrete transmission levels arise as the transition
threshold is reached for different detunings in each domain.

In order to simulate the non-equilibrium behavior of the spatial domains
in our phenomenological model, we consider distinct fractions $f_{R1}$
and $f_{R2}$ to each domain and their overall optical response.
We then plot a color map of the transmission difference for the $+/-$
scan directions against $\text{\ensuremath{f_{\text{R2}}}}$ and $\Delta_{c}$
by setting $f_{\text{r1}}=0.33$, as in Fig.~\ref{multibistability-tuning}(a).
For the experiments, we extract the transmission difference for the
$+/-$ scan directions and plot the probe transmission versus $\Omega_{\textrm{pert}}$
and $\Delta_{c}$ in Fig.~\ref{multibistability-tuning}(b). In both
theory and experiments, we observe two bistable regimes, and a multistable
regime with overlapping regions. We also plot the transmission for the $+/-$ scan directions for theory and experiments, as
shown in Figs.~\ref{multibistability-tuning}(c) and (d). The
optical responses for multiple spatial domains show a double
optical bistability, as described in \citep{supplement}.

It is insightful to increase the number of spatial domains
in the experiments, and hence the number of phase jumps, by adding a
few perturbing laser beams. Figures~\ref{multi stability}(a-d)
show the transmission spectra with 1, 2, and 3 discrete jumps, obtained by
scanning $\Delta_{c}$ from red- to blue-detuning.
Figure~\ref{multi stability}(a) refers to the single domain created by the probe field,
showing a sudden jump in the transmission of light
[gray region (1), where  $ \Delta_{c} / 2\pi \sim -19 $ MHz].
As we add the perturbing field 1 overlapping with the probe,
the jump in Fig.~\ref{multi stability}(a) is split into two jumps
in Fig.~\ref{multi stability}(b), these located at the detuning
$\Delta_{c}/2\pi\sim-19$ MHz and $\Delta_{c}/2\pi\sim-24$ MHz, respectively.
The three jumps in Fig.~\ref{multi stability}(d) correspond to
three regions of higher Rydberg density associated with the probe
and the perturbing fields 1 and 2.
This added interplay between multiple spatial domains results in additional
complexity that manifests in the additional jumps in Fig.~\ref{multi stability}(b).
Meanwhile, the phase transition is weakened, as seen from the decreased
height of each jump and the susceptibility of the phase transition
defined as $d\textrm{T}/d\Delta_{c}$ ($d\textrm{T}$ is the transmission difference of the jump)
is reduced near the transition point $\Delta_{c}/2\pi\sim-19$ MHz.
Figure~\ref{multi stability}(c) shows the corresponding situation
when applying a perturbing field 2. In these two cases, as $\Delta_{c}$
approaches resonance, $f_{R}$ increases and eventually one domain
reaches the threshold $f_R^{(1)}$ at position 1 and another domain
reaches the threshold $f_R^{(2)}$, the system undergoes two non-equilibrium
phase transitions accompanied by two sudden jumps in the optical transmission.
Figure~\ref{multi stability}(d) shows three jumps created by both
the probe and perturbing fields 1 and 2. Changing the alignment
determines whether these regions overlap or are separate. Here, we
extend the epidemic model to multiple spatial domains according to
the 2D SIS cellular automaton. The qualitative agreement between
the predicted phase jumps, Figs.~\ref{multi stability}(e-h), and
the experimental observations, Figs.~\ref{multi stability}(a-d),
confirms that we can associate each jump with an individual domain.

Non-equilibrium phase transitions within multi-domains often correspond
to optical multistability, as demonstrated here, which is the extension
of optical bistability to three or more stable output states. Before
this work, multistability has been predicted and investigated in coupled
atom-cavity systems \citep{kitano1981optical,cecchi1982observation,joshi2003optical,sheng2012realization}
and semiconductor microcavities \citep{gippius2007polarization,paraiso2010multistability,sarkar2010polarization,cerna2013ultrafast,goblot2019nonlinear}.
Optical bistability in Rydberg atoms has previously been studied both in theory \citep{lee2012collective,marcuzzi2014universal,weimer2015variational,vsibalic2016driven,levi2016quantum,vsibalic2016driven} and experiments \citep{carr2013nonequilibrium,malossi2014full,de2016intrinsic,weller2016charge,ding2020phase}. However, optical multistability in Rydberg atoms is more challenging, and has not been explored before.
A mean-field master Bloch equation can be used to
qualitatively simulate multistability, because the individual
Rydberg atoms interact with each other at random, in such a way that
each Rydberg atom in a compartment is treated similarly and
indistinguishably. More precise simulations in large-scale networks must
consider the subtle differences between various states of the Rydberg atoms.

In summary, we have studied optically-driven multi-domain non-equilibrium dynamics in strongly interacting Rydberg atom gases. The measured hysteresis and phase jumps can be understood well and reproduced qualitatively using the epidemic model and using a mean-field master equation. Our results highlight the rich range of non-equilibrium phenomena that are accessible even in a relatively simple experiment and provide observational data to benchmark theoretical models of non-equilibrium dynamics under arbitrary spatial structures. Specifically, the observed dynamics of Rydberg atoms in different time and space scales could predict ``herd immunity", ``endemic state" and the outbreak of a disease in multiple location. The reported multi-domain
dependent exotic phases could help build a Rydberg-based simulate platform for more complex phenomena in real-world, such as in epidemiology \citep{pastor2015epidemic}, ecosystems and climate \citep{haken2006information} or other complex systems \citep{morgado2020quantum}.

\begin{acknowledgments}
We thank for Christopher G. Wade, Kevin J. Weatherill and Igor Lesanovsky for helpful discussions on the phase boundaries and non-equilibrium dynamics. We acknowledge funding from National Key R\&D Program of China (2017YFA0304800), NSFC funding (Grant Nos. U20A20218, 61525504, 61722510, 61435011), the Youth Innovation Promotion Association of CAS Grant No. 2018490, EPSRC through grant agreements EP/M014398/1, EP/R002061/1, EP/L023024/1, EP/P012000/1, EP/R035482/1, EP/S015973/1, as well as, DSTL, and Durham University. The European Union's Horizon 2020 Research and Innovation Program under Grant No. 845218 (Marie Curie Fellowship to H. B.). F.N. was supported in part by NTT Research, JSPS Grant No. JP20H00134, ARO Grant No. W911NF-18-1-0358, AOARD Grant No. FA2386-20-1-4069, and FQXi Grant No. FQXi-IAF19-06.
\end{acknowledgments}

\bibliographystyle{ScienceAdvances}

\begin{thebibliography}{10}

\bibitem{haken2006information}
H.~Haken, {\it Information and self-organization: A macroscopic approach to
  complex systems\/} (Springer Science \& Business Media, 2006).

\bibitem{per1987self}
P.~Bak, C.~Tang, K.~Wiesenfeld, Self-organized criticality: and explanation of 1/f
  noise.
\newblock {\it Phys. Rev. Lett.\/} {\bf 59}, 381--384 (1987).

\bibitem{field1995superconducting}
S.~Field, J.~Witt, F.~Nori, X.~Ling, Superconducting vortex avalanches.
\newblock {\it Phys. Rev. Lett.\/} {\bf 74}, 1206 (1995).

\bibitem{olson1997superconducting}
C.~Olson, C.~Reichhardt, F.~Nori, Superconducting vortex avalanches, voltage
  bursts, and vortex plastic flow.
  %: Effect of the microscopic pinning landscape on the macroscopic properties.
\newblock {\it Phys. Rev. B.\/} {\bf 56}, 6175 (1997).

\bibitem{schauss2012observation}
P.~Schau{\ss}, {\it et~al.\/}, Observation of spatially ordered structures in a
  two-dimensional {Rydberg} gas.
\newblock {\it Nature\/} {\bf 491}, 87--91 (2012).

\bibitem{labuhn2016tunable}
H.~Labuhn, {\it et~al.\/}, Tunable two-dimensional arrays of single {Rydberg}
  atoms for realizing quantum {Ising} models.
\newblock {\it Nature\/} {\bf 534}, 667--684 (2016).

\bibitem{bernien2017probing}
H.~Bernien, {\it et~al.\/}, Probing many-body dynamics on a 51-atom quantum
  simulator.
\newblock {\it Nature\/} {\bf 551}, 579--584 (2017).

\bibitem{de2019observation}
S.~de~L{\'e}s{\'e}leuc, {\it et~al.\/}, Observation of a symmetry-protected
  topological phase of interacting bosons with {Rydberg} atoms.
\newblock {\it Science\/} {\bf 365}, 775--780 (2019).

\bibitem{bluvstein2021controlling}
D.~Bluvstein, {\it et~al.\/}, Controlling quantum many-body dynamics in driven
  Rydberg atom arrays.
\newblock {\it Science\/} {\bf 371}, 1355--1359 (2021).

\bibitem{eisert2015quantum}
J.~Eisert, M.~Friesdorf, C.~Gogolin, Quantum many-body systems out of
  equilibrium.
\newblock {\it Nature Physics\/} {\bf 11}, 124--130 (2015).

\bibitem{browaeys2020many}
A.~Browaeys, T.~Lahaye, Many-body physics with individually controlled {Rydberg}
  atoms.
\newblock {\it Nature Physics\/} {\bf 16}, 132--142 (2020).

\bibitem{saffman2010quantum}
M.~Saffman, T.~Walker, K.~M{\o}lmer, Quantum information with {Rydberg} atoms.
\newblock {\it Reviews of Modern Physics\/} {\bf 82}, 2313 (2010).

\bibitem{Adams_2019}
C.~S. Adams, J.~D. Pritchard, J.~P. Shaffer, Rydberg atom quantum technologies.
\newblock {\it Journal of Physics B\/}
  {\bf 53}, 012002 (2019).

\bibitem{morgado2020quantum}
M.~Morgado, S.~Whitlock, Quantum simulation and computing with
  {Rydberg}-interacting qubits.
\newblock {\it AVS Quantum Sci\/} {\bf 3}, 023501 (2021).

\bibitem{bloch2012quantum}
I.~Bloch, J.~Dalibard, S.~Nascimbene, Quantum simulations with ultracold
  quantum gases.
\newblock {\it Nature Physics\/} {\bf 8}, 267--276 (2012).

\bibitem{georgescu2014quantum}
I.~M. Georgescu, S.~Ashhab, F.~Nori, Quantum simulation.
\newblock {\it Reviews of Modern Physics\/} {\bf 86}, 153 (2014).

\bibitem{gallagher2005rydberg}
T.~F. Gallagher, {\it Rydberg atoms\/}, vol.~3 (Cambridge University Press,
  2005).

\bibitem{firstenberg2016nonlinear}
O.~Firstenberg, C.~S. Adams, S.~Hofferberth, Nonlinear quantum optics mediated
  by {Rydberg} interactions.
\newblock {\it Journal of Physics B\/}
  {\bf 49}, 152003 (2016).

\bibitem{urvoy2015strongly}
A.~Urvoy, {\it et~al.\/}, Strongly correlated growth of {Rydberg} aggregates in
  a vapor cell.
\newblock {\it Phys. Rev. Lett.\/} {\bf 114}, 203002 (2015).

\bibitem{carr2013nonequilibrium}
C.~Carr, R.~Ritter, C.~Wade, C.~S. Adams, K.~J. Weatherill, Nonequilibrium
  phase transition in a dilute {Rydberg} ensemble.
\newblock {\it Phys. Rev. Lett.\/} {\bf 111}, 113901 (2013).

\bibitem{malossi2014full}
N.~Malossi, {\it et~al.\/}, Full counting statistics and phase diagram of a
  dissipative {Rydberg} gas.
\newblock {\it Phys. Rev. Lett.\/} {\bf 113}, 023006 (2014).

\bibitem{marcuzzi2014universal}
M.~Marcuzzi, E.~Levi, S.~Diehl, J.~P. Garrahan, I.~Lesanovsky, Universal
  nonequilibrium properties of dissipative {Rydberg} gases.
\newblock {\it Phys. Rev. Lett.\/} {\bf 113}, 210401 (2014).

\bibitem{helmrich2020signatures}
S.~Helmrich, {\it et~al.\/}, Signatures of self-organized criticality in an
  ultracold atomic gas.
\newblock {\it Nature\/} {\bf 577}, 481--486 (2020).

\bibitem{ding2020phase}
D.-S. Ding, H.~Busche, B.-S. Shi, G.-C. Guo, C.~S. Adams, Phase diagram and
  self-organizing dynamics in a thermal ensemble of strongly interacting
  {Rydberg} atoms.
\newblock {\it Phys. Rev. X.\/} {\bf 10}, 021023 (2020).

\bibitem{wintermantel2021epidemic}
T.~Wintermantel, {\it et~al.\/}, Epidemic growth and {G}riffiths effects on an
  emergent network of excited atoms.
\newblock {\it Nature Communications\/} {\bf 12}, 1--6 (2021).

\bibitem{klocke2021hydrodynamic}
K.~Klocke, T.~Wintermantel, G.~Lochead, S.~Whitlock, M.~Buchhold, Hydrodynamic
  stabilization of self-organized criticality in a driven {Rydberg} gas.
\newblock {\it Phys. Rev. Lett.\/} {\bf 126}, 123401 (2021).

\bibitem{eubank2004modelling}
S.~Eubank, {\it et~al.\/}, Modelling disease outbreaks in realistic urban
  social networks.
\newblock {\it Nature\/} {\bf 429}, 180--184 (2004).

\bibitem{longini2005containing}
I.~M. Longini, {\it et~al.\/}, Containing pandemic influenza at the source.
\newblock {\it Science\/} {\bf 309}, 1083--1087 (2005).

\bibitem{pastor2015epidemic}
R.~Pastor-Satorras, C.~Castellano, P.~Van~Mieghem, A.~Vespignani, Epidemic
  processes in complex networks.
\newblock {\it Reviews of modern physics\/} {\bf 87}, 925 (2015).

\bibitem{supplement}
Supplementary materials, including experimental setup, optical multistability,
  theoretical analysis and fast and slow scan effect.

\bibitem{drossel1992self}
B.~Drossel, F.~Schwabl, Self-organized critical forest-fire model.
\newblock {\it Phys. Rev. Lett.\/} {\bf 69}, 1629 (1992).

\bibitem{kitano1981optical}
M.~Kitano, T.~Yabuzaki, T.~Ogawa, Optical tristability.
\newblock {\it Phys. Rev. Lett.\/} {\bf 46}, 926 (1981).

\bibitem{cecchi1982observation}
S.~Cecchi, G.~Giusfredi, E.~Petriella, P.~Salieri, Observation of optical
  tristability in sodium vapors.
\newblock {\it Phys. Rev. Lett.\/} {\bf 49}, 1928 (1982).

\bibitem{joshi2003optical}
A.~Joshi, M.~Xiao, Optical multistability in three-level atoms inside an
  optical ring cavity.
\newblock {\it Phys. Rev. Lett.\/} {\bf 91}, 143904 (2003).

\bibitem{sheng2012realization}
J.~Sheng, U.~Khadka, M.~Xiao, Realization of all-optical multistate switching
  in an atomic coherent medium.
\newblock {\it Phys. Rev. Lett.\/} {\bf 109}, 223906 (2012).

\bibitem{gippius2007polarization}
N.~Gippius, {\it et~al.\/}, Polarization multistability of cavity polaritons.
\newblock {\it Phys. Rev. Lett.\/} {\bf 98}, 236401 (2007).

\bibitem{paraiso2010multistability}
T.~Para{\"\i}so, M.~Wouters, Y.~L{\'e}ger, F.~Morier-Genoud,
  B.~Deveaud-Pl{\'e}dran, Multistability of a coherent spin ensemble in a
  semiconductor microcavity.
\newblock {\it Nature Materials\/} {\bf 9}, 655--660 (2010).

\bibitem{sarkar2010polarization}
D.~Sarkar, {\it et~al.\/}, Polarization bistability and resultant spin rings in
  semiconductor microcavities.
\newblock {\it Phys. Rev. Lett.\/} {\bf 105}, 216402 (2010).

\bibitem{cerna2013ultrafast}
R.~Cerna, {\it et~al.\/}, Ultrafast tristable spin memory of a coherent
  polariton gas.
\newblock {\it Nature Communications\/} {\bf 4}, 2008 (2013).

\bibitem{goblot2019nonlinear}
V.~Goblot, {\it et~al.\/}, Nonlinear polariton fluids in a flatband reveal
  discrete gap solitons.
\newblock {\it Phys. Rev. Lett.\/} {\bf 123}, 113901 (2019).

\bibitem{lee2012collective}
T.~E. Lee, H.~Haeffner, M.~Cross, Collective quantum jumps of {Rydberg} atoms.
\newblock {\it Phys. Rev. Lett.\/} {\bf 108}, 023602 (2012).

\bibitem{weimer2015variational}
H.~Weimer, Variational principle for steady states of dissipative quantum
  many-body systems.
\newblock {\it Phys. Rev. Lett.\/} {\bf 114}, 040402 (2015).

\bibitem{vsibalic2016driven}
N.~{\v{S}}ibali{\'c}, C.~G. Wade, C.~S. Adams, K.~J. Weatherill, T.~Pohl,
  Driven-dissipative many-body systems with mixed power-law interactions:
  Bistabilities and temperature-driven nonequilibrium phase transitions.
\newblock {\it Phys. Rev. A.\/} {\bf 94}, 011401 (2016).

\bibitem{levi2016quantum}
E.~Levi, R.~Guti{\'e}rrez, I.~Lesanovsky, Quantum non-equilibrium dynamics of
  {Rydberg} gases in the presence of dephasing noise of different strengths.
\newblock {\it Journal of Physics B\/}
  {\bf 49}, 184003 (2016).

\bibitem{de2016intrinsic}
N.~R. de~Melo, {\it et~al.\/}, Intrinsic optical bistability in a strongly
  driven {Rydberg} ensemble.
\newblock {\it Phys. Rev. A.\/} {\bf 93}, 063863 (2016).

\bibitem{weller2016charge}
D.~Weller, A.~Urvoy, A.~Rico, R.~L{\"o}w, H.~K{\"u}bler, Charge-induced optical
  bistability in thermal {Rydberg} vapor.
\newblock {\it Phys. Rev. A.\/} {\bf 94}, 063820 (2016).

% \bibitem{grafke2017spatiotemporal}
% T.~Grafke, M.~E. Cates, E.~Vanden-Eijnden, Spatiotemporal self-organization of
%   fluctuating bacterial colonies.
% \newblock {\it Phys. Rev. Lett.\/} {\bf 119}, 188003 (2017).

%\bibitem{majdandzic2014spontaneous}
%A.~Majdandzic, {\it et~al.\/}, Spontaneous recovery in %dynamical networks.
%\newblock {\it Nature Physics\/} {\bf 10}, 34 (2014).

% \bibitem{weller2019interplay}
% D.~Weller, J.~P. Shaffer, T.~Pfau, R.~L{\"o}w, H.~K{\"u}bler, Interplay between
%  thermal {Rydberg} gases and plasmas.
% \newblock {\it Phys. Rev. A.\/} {\bf 99}, 043418 (2019).

% \bibitem{wade2018terahertz}
% C.~G. Wade, {\it et~al.\/}, A terahertz-driven non-equilibrium phase transition
%   in a room temperature atomic vapour.
% \newblock {\it Nature Communications\/} {\bf 9}, 3567 (2018).

\end{thebibliography}

\end{document}

\title{Supplementary materials for: Epidemic spreading and \\
herd immunity in a driven non-equilibrium system\\
of strongly-interacting atoms}
\author{Dong-Sheng Ding}
\email{dds@ustc.edu.cn}

\affiliation{Key Laboratory of Quantum Information, University of Science and Technology
of China, Hefei, Anhui 230026, China.}
\affiliation{Synergetic Innovation Center of Quantum Information and Quantum Physics,
University of Science and Technology of China, Hefei, Anhui 230026,
China.}
\author{Zong-Kai Liu}
\affiliation{Key Laboratory of Quantum Information, University of Science and Technology
of China, Hefei, Anhui 230026, China.}
\affiliation{Synergetic Innovation Center of Quantum Information and Quantum Physics,
University of Science and Technology of China, Hefei, Anhui 230026,
China.}
\author{Hannes Busche}
\affiliation{Department of Physics, Chemistry and Pharmacy, Physics@SDU, University
of Southern Denmark, 5230 Odense M, Denmark }
\author{Bao-Sen Shi}
\email{drshi@ustc.edu.cn}

\affiliation{Key Laboratory of Quantum Information, University of Science and Technology
of China, Hefei, Anhui 230026, China.}
\affiliation{Synergetic Innovation Center of Quantum Information and Quantum Physics,
University of Science and Technology of China, Hefei, Anhui 230026,
China.}
\author{Guang-Can Guo}
\affiliation{Key Laboratory of Quantum Information, University of Science and Technology
of China, Hefei, Anhui 230026, China.}
\affiliation{Synergetic Innovation Center of Quantum Information and Quantum Physics,
University of Science and Technology of China, Hefei, Anhui 230026,
China.}
\author{Charles S. Adams}
\email{c.s.adams@durham.ac.uk}

\affiliation{Department of Physics, Joint Quantum Centre (JQC) Durham-Newcastle,
Durham University, South Road, Durham DH1 3LE, United Kingdom}
\author{Franco Nori}
\email{fnori@riken.jp}

\affiliation{Theoretical Quantum Physics Laboratory, RIKEN Cluster for Pioneering
Research, Wako-shi, Saitama 351-0198, Japan.}
\affiliation{RIKEN Center for Quantum Computing (RQC), Wako-shi, Saitama 351-0198,
Japan.}
\affiliation{Physics Department, The University of Michigan, Ann Arbor, MI 48109-1040,
USA.}
\date{\today}

\maketitle

\textbf{I. Experimental setup}

\begin{figure}[h]
\includegraphics[width=1\columnwidth]{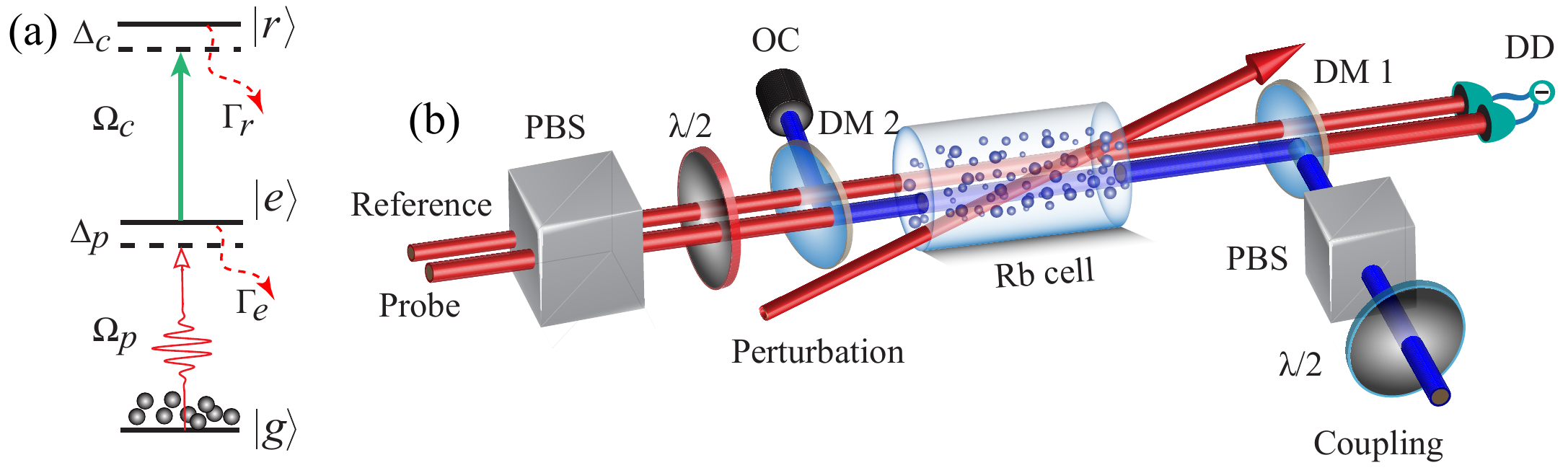}\caption{(a) Energy diagram used to excite Rydberg atoms via a two-photon transition using probe and coupling fields with Rabi frequencies, $\Omega_{p}$ and $\Omega_{c}$, respectively. (b) Overview of the experimental setup. The probe (and an identical reference beam resonant with the $|g\rangle=|5S_{1/2},F=3\rangle\rightarrow|e\rangle=|5P_{1/2},F'=2\rangle$ transition in $^{85}$Rb) propagate in parallel through a heated Rb cell. The coupling light (blue), tuned close to resonance with the $|e\rangle\rightarrow|r\rangle=|47D_{3/2}\rangle$ transition, counter-propagates along the axis of the probe. Additional perturbing fields, with Rabi frequencies $\Omega_{\textrm{pert}}$, are applied at an angle $\theta$. The probe and reference transmission are detected on a differencing photodetector (DD). PBS: polarizing beam splitter, DM: dichroic mirror, $\lambda/2$: half-wave plate, DD: differencing photodetector, OC: optical collector.}
\label{Experimental setup-1}
\end{figure}

The energy diagram and experimental set-up are schematically shown in Fig.~\ref{Experimental setup-1}(a)
and (b), respectively. Aside from the addition of the perturbing field,
the setup is similar to the one in Ref.~\cite{ding2019Phase}. Three atomic
levels, a ground state $\lvert g\rangle$, a low-lying excited state
$\lvert e\rangle$, and a Rydberg state $\lvert r\rangle$, are driven
by a probe (near-resonant with $\lvert g\rangle\rightarrow\lvert e\rangle$)
and a coupling field (near-resonant with $\lvert e\rangle\rightarrow\lvert r\rangle$).
These fields counter-propagate through a $5\,\mathrm{cm}$ long vapour
cell filled with $^{85}$Rb atoms at $T=60^{\circ}\mathrm{C}$. The
two lasers excite Rydberg atoms in a ladder-type electromagnetically
induced transparency (EIT) configuration \cite{mohapatra2007coherent}.
The ground state atom density is $3.3\times10^{11}\,\mathrm{cm}^{-3}$,
corresponding to a mean interatomic spacing of $0.8\,\mathrm{\mu m}$.
The probe beam is focused into the cell ($1/e^{2}$-waist radius $\sim$ $500~\mu$m) and couples the ground state $\lvert g\rangle=\lvert5S_{1/2},F=3\rangle$
to $\lvert e\rangle=\lvert5P_{1/2},F'=2\rangle$ with detuning $\Delta_{p}$
and Rabi frequency $\Omega_{p}$. The coupling beam ($1/e^{2}$-waist
radius of $\sim$ $200\,\mathrm{\mu m}$) with detuning $\Delta_{c}$ is resonant
with the transition from $\lvert e\rangle$ to a Rydberg state $\lvert r\rangle=\lvert47D_{3/2}\rangle$
with Rabi frequency $\Omega_{c}\sim2\pi\times20$ MHz.
By using a counter-propagating geometry, the Doppler effect is reduced
by a factor of $v(w_{p}-w_{c})/c$, where $v$ is an atom's velocity,
and $\omega_{p}$ and $\omega_{c}$ denote the angular frequencies
of the probe and coupling light. If the lifetime of $\lvert r\rangle$
is long compared to $\lvert e\rangle$, (i.e. the decay rates satisfy
$\Gamma_{r}\ll\Gamma_{e}$, as typically the case for Rydberg states)
and the probe is weak compared to the coupling field $\Omega_{p}\ll\Omega_{c}$,
this corresponds to an EIT configuration where the presence of the
resonant coupling field renders the ensemble transparent for the resonant
probe light. In the experiment, we fix $\Delta_{p}=0$ and observe
the transmission as $\Delta_{c}$ is scanned at different rates.
The transmission of a reference field (sent through the cell in parallel
and not overlapping with the coupling light) is subtracted via a pair
of balanced amplified photodiodes. The photoelectric signal of the transmission difference between probe and reference is recorded by a computer.

The additional perturbation schematically shown in Fig.~\ref{Experimental setup-1}(b) represents either one or two perturbing fields, which are of the same type as the probe beam. The Rabi frequencies and angles of these perturbing fields could be tuned individually, to create a spatially inhomogeneous
Rydberg excitation rate. In addition, the perturbing fields are coupled to the coupling field in order to satisfy the EIT configuration. They only modulate the Rydberg populations along the path of the probe field, but are not collected by the balanced amplified photodiodes. In this case, the non-equilibrium dynamics of the localized system along the path of the probe field is strongly affected by the perturbation.

\clearpage{}

\textbf{II. Optical multistability}

\begin{figure}[h]
\includegraphics[width=1\columnwidth]{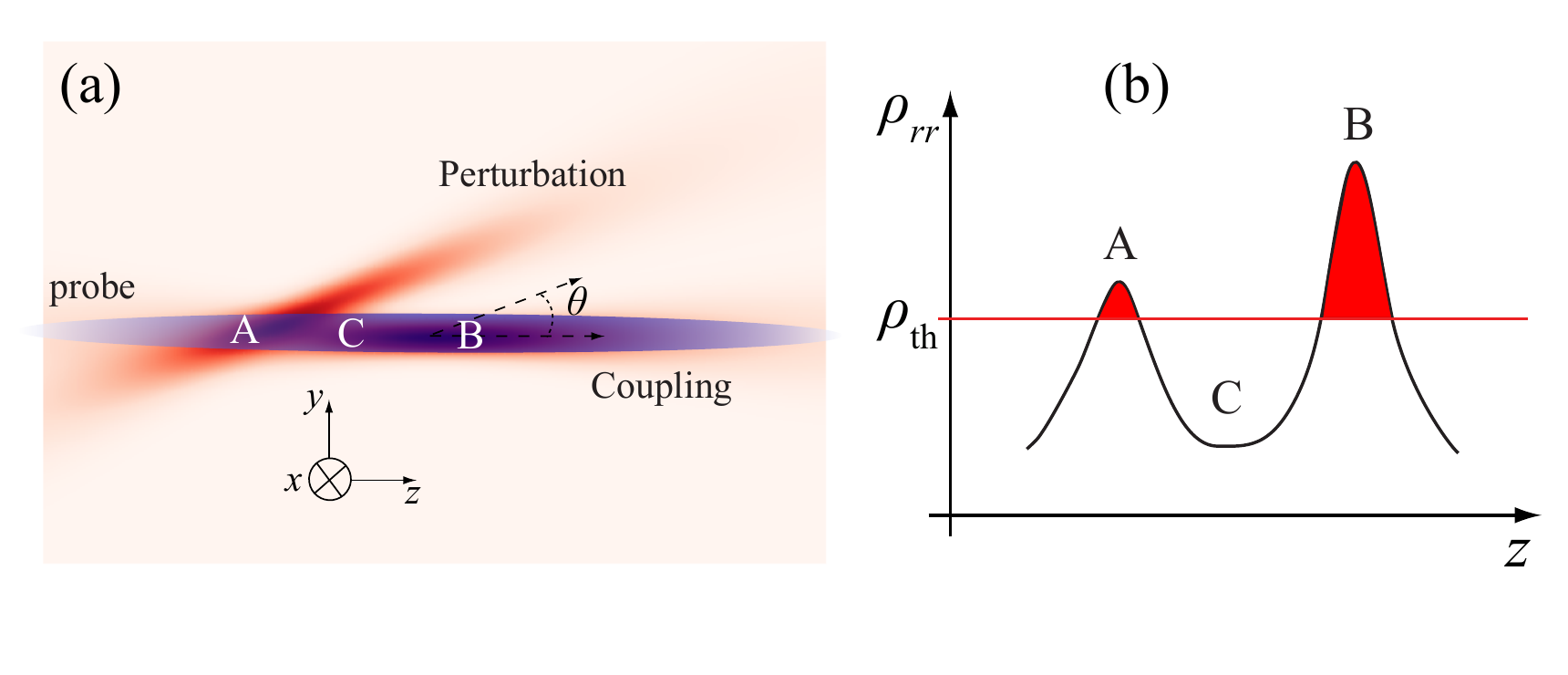}\caption{(a) Schematic diagram illustrating the intensity distribution inside
the cell. The perturbing beam and focused probe form two distinct
high intensity regions labelled A and B, respectively. The relative
Rydberg populations in A and B can be controlled by varying $\Omega_{\textrm{pert}}$.
(b) Schematic illustration of the Rydberg atom density $\rho_{rr}$
along the optical axis $z$-coordinate. When $\rho_{rr}>\rho_{\textrm{th}}$,
the critical Rydberg density, a phase transition between a non-interacting
and the interacting phase occurs. In region C, between the two domains
A and B, the Rydberg density is below threshold.}
\label{density distribution}
\end{figure}

A dynamical phase transition can be observed as a jump in the probe transmission as the critical Rydberg density [corresponding to the critical population $\rho_{\textrm{th}}$, shown as a horizontal red line in Fig.~\ref{density distribution}(b)] is exceeded, and an optical bistability window can be observed depending on the scan direction and rate of change of $\Delta_{c}$. The areas beyond the threshold $\rho_{\textrm{th}}$ are marked in red in Fig.~\ref{density distribution} (b). To demonstrate how spatial domains\textemdash with different excitation dynamics\textemdash result in optical multistability, we expose part of the excitation volume to an additional probe light in the form of a single perturbing beam that intersects the probe and control beams at an angle $\theta$, and locally enhances the Rydberg population in the region labelled A in Fig.~\ref{Experimental setup-1}(a), also marked as A in Fig.~\ref{Experimental setup-1}(b). A second local maximum labelled B in Fig.~\ref{Experimental setup-1}(a) occurs at the focus of the probe.

The amplitude of the perturbing field is characterized by a Rabi frequency
$\Omega_{\textrm{pert}}$, which can be varied independently of $\Omega_{p}$.
Varying $\Omega_{\textrm{pert}}$ can adjust the
relative intensity and hence the Rydberg density in regions A and
B. As the two domains are characterized by distinct Rydberg populations,
they can exhibit either independent or coupled non-equilibrium dynamics
depending on the size of the separation region labelled C in Fig.~\ref{density distribution}. Note that a domain boundary can also arise from attenuating the probe light along the medium \citep{carr2013nonequilibrium}. The dynamics of the optical multistability along the $z$-axis is thus affected by the overall inhomogeneity of the Rydberg density. For any spatially
varying intensity distribution, we expect distinct phases that are
characterized by a spatial boundary between states of high and low
population \citep{inguva1990spatial}. In addition to the control
offered via $\Omega_{\textrm{pert}}$, we can also vary both the total
intensity in region A and its overlap with region B via the alignment
of the perturbing beam. As expected, we find that both the intensity
in region A and the separation to region B, i.e. C in Fig.~\ref{density distribution} are critical. Additional perturbing fields would create more domains.

\clearpage{}

\textbf{III. Theoretical analysis}

Here we show that the experimental results can be modeled using mean-field
optical Bloch equations and qualitatively using an epidemic model.
The Lindblad master equation is given by
\begin{equation}
d\rho/dt=-i\left[H,\rho\right]/\hbar+L/\hbar,
\end{equation}
where $\rho$ is the atomic ensemble's density matrix and $H=\sum_{k}H[\rho^{(k)}]$
the atom-light interaction Hamiltonian summed over all the single-atom
Hamiltonians using the rotating wave approximation. The single-atom
Hamiltonian has the form:
\begin{equation}
H[\rho^{(k)}]=-\frac{\hbar}{2}\begin{pmatrix}0 & \Omega_{p} & 0\\
\Omega_{p} & -2\Delta_{p} & \Omega_{c}\\
0 & \Omega_{c} & -2(\Delta_{p}-\Delta_{c})
\end{pmatrix}
\end{equation}
The Lindblad superoperator $L=\sum_{k}L[\rho^{(k)}]$ is comprised
of the single-atom superoperators, where $L[\rho^{(k)}]/\hbar=-\frac{1}{2}\sum_{m}\left(C_{m}^{\dagger}C_{m}\rho+\rho C_{m}^{\dagger}C_{m}\right)+\sum_{m}C_{m}\rho C_{m}^{\dagger}$, $C_{1}=\sqrt{\Gamma_{e}}\left|g\right\rangle \left\langle e\right|$,
and $C_{2}=\sqrt{\Gamma_{r}}\left|e\right\rangle \left\langle r\right|$
are collapse operators for the decay from state $\left|e\right\rangle $
to $\left|g\right\rangle $ and from $\left|r\right\rangle $ to $\left|e\right\rangle $
with rates $\Gamma_{e}$ and $\Gamma_{r}$ respectively. The matrix
form of the Lindbladian can be written as

\begin{equation}
\frac{L[\rho^{(k)}]}{\hbar}=\frac{1}{2}\left(\begin{array}{ccc}
2\Gamma_{e}\rho_{ee}^{(k)} & -\Gamma_{e}\rho_{ge}^{(k)} & -\Gamma_{r}\rho_{gr}^{(k)}\\
-\Gamma_{e}\rho_{eg}^{(k)} & -2\Gamma_{e}\rho_{ee}^{(k)}+2\Gamma_{r}\rho_{rr}^{(k)} & -\left(\Gamma_{e}+\Gamma_{r}\right)\rho_{er}^{(k)}\\
-\Gamma_{r}\rho_{rg}^{(k)} & -\left(\Gamma_{e}+\Gamma_{r}\right)\rho_{re}^{(k)} & -2\Gamma_{r}\rho_{rr}^{(k)}
\end{array}\right)
\end{equation}

\begin{figure}[h]
\includegraphics[width=1\columnwidth]{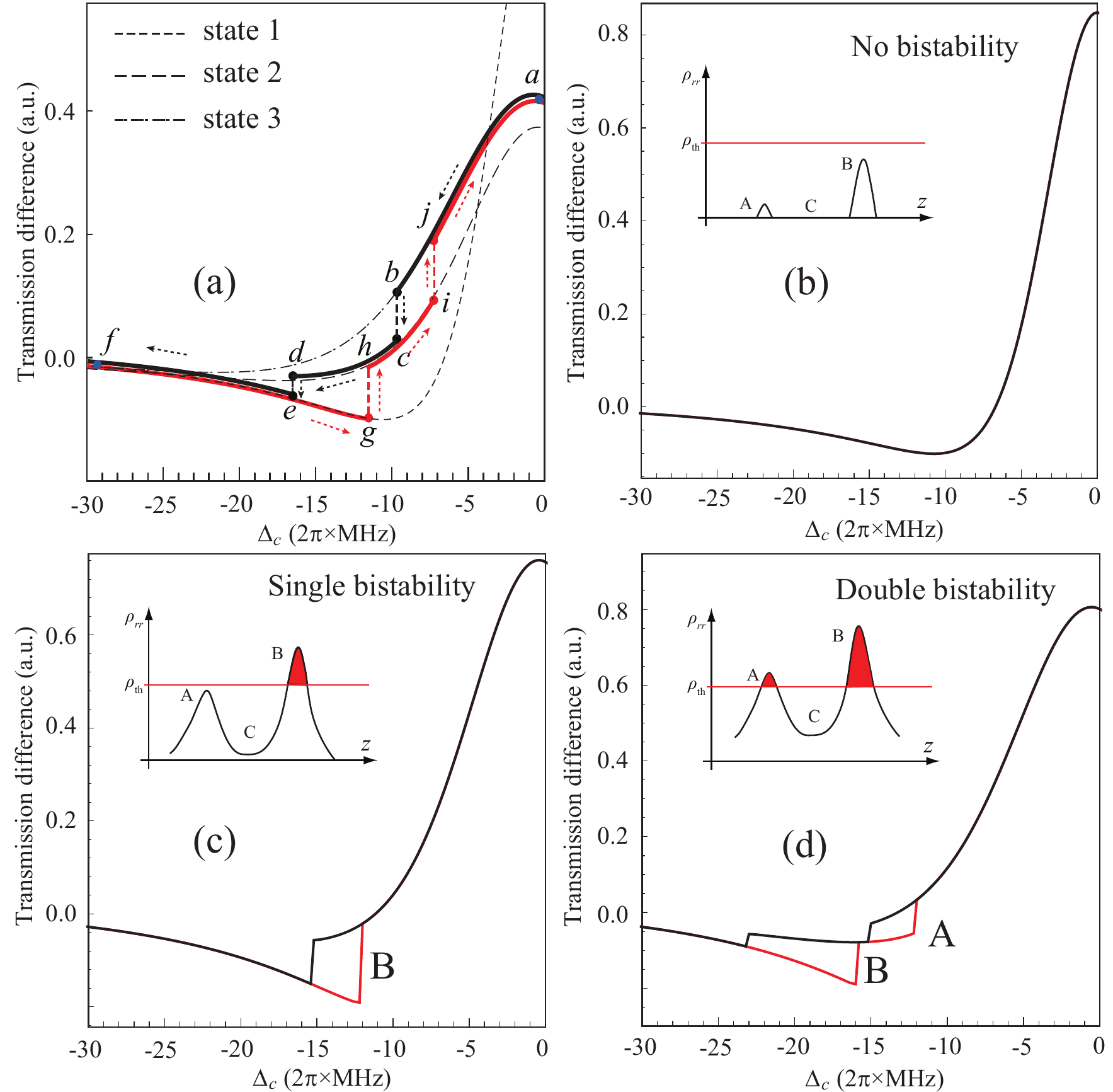}\caption{(a) Simplified theoretical simulation of multi-domain bistability.
The gray dotted lines correspond to steady states 1, 2 and 3, respectively.
The red and black lines are the trajectories when scanning from red-
to blue-detuning (red) and vice versa (black). State 1 has a normal
EIT transmission with $T(0<f_{R}<0.09)$. State 2 shows the integrated
transmission $T(f_{R}=0.23)$. State 3 corresponds to $T(f_{R}=0.33)$.
(b-d) Three cases are compared for (b) no, (c) single bistability
in region B only and (d) double bistability in regions A and B. The
inserted subfigures describe the Rydberg population under three situations: (b)
both of regions A and B are below $\rho_{\textrm{th}}$; (c) Region A is below and B is above $\rho_{\textrm{th}}$;
(d) both of regions A and B are above $\rho_{\textrm{th}}$.}
\label{transmission}
\end{figure}

Since we are only concerned with the steady state, i.e. $t\rightarrow\infty$,
the Lindblad master equation can be solved for $d\rho/dt=0$. The
complex susceptibility of the EIT medium including the Doppler effect
due to atomic motion is
\begin{equation}
\chi(v)\mathrm{d}v=(\lvert\mu_{ge}\rvert^{2}/\epsilon_{0}\hbar)\rho_{eg}(v)\mathrm{d}v,
\end{equation}
the coherence $\rho_{eg}$ in the density matrix is obtained by solving
the master equation. It has the form
\begin{equation}
\rho_{eg}(v)=\frac{N(v)[\Gamma_{r}+2i(\delta+\Delta_{D})]}{(2\Delta_{p}+2\omega_{p}v/c-i\Gamma_{e})[\Gamma_{r}+2i(\delta+\Delta_{D})-i\Omega_{c}^{2}]}.
\end{equation}
where $\Delta_{D}=(\omega_{p}-\omega_{c})v/c$ denotes the Doppler shift
experienced by an atom moving with velocity $v$, $\delta=\Delta_{c}+\Delta_{p}$ is
the two-photon detuning, and $N(v)$ is the effective atom number.
The spectra of the EIT medium are obtained from the susceptibility
via
\begin{equation}
T(\Delta_{c},\Gamma_{r})\sim \exp\{-{\rm Im}[\int kL\chi(v)dv]\}.
\end{equation}
where $L$ is the medium length and $k$ the wavevector of the probe
field. In the Doppler integration, we consider atomic velocities from
$v=-500$ m/s to $500$ m/s in order to reduce computational complexity.
Due to the two spatial domains A and B illustrated in the inset of
Fig.~\ref{density distribution}(b), we assume that there are three
distinct steady states, 1, 2, and 3, corresponding to: no domains above
threshold; one domain above threshold; and two domains above threshold,
respectively. The steady-state lineshapes associated with these three
steady-states are plotted as dashed lines in Fig.~\ref{transmission}.
To calculate these steady-state lineshapes, we assume that both the
energy and the decay rate, $\Gamma_{r}$, of the Rydberg state are
modified by motional averaging of interaction potentials and Stark
shifts $\Delta_{\textrm{shift}}$ (due to electric fields created by ions)
\cite{carr2013nonequilibrium,weller2016charge}. These effects are
described by
\begin{align}
\Delta_{c} & \to\Delta_{c}+\eta_{1}\cdot(f_{R}-f_{R,c}),
\end{align}
and
\begin{align}
\Gamma_{r} & \to\Gamma_{r}+\eta_{2}\cdot(f_{R}-f_{R,c}),
\end{align}
when $f_{R}>f_{R,c}$, where $f_{R}$ is the fraction of Rydberg atoms,
and we set the critical Rydberg fraction, $f_{R,c}=0.09$, $\eta_{1}=3$
MHz and $\eta_{2}=50$ MHz to match the experimental results. The
observation that the broadening exceeds the line shift ($\eta_{2}>\eta_{1}$)
is consistent with previous work on Rydberg-induced bistability \cite{weller2016charge,ding2019Phase}.

When $\Delta_{c}$ is scanned from red- to blue-detuning (positive
+ scan direction, red line), the transmission initially evolves along
the normal EIT resonance curve (steady state 1). As $\Delta_{c}$
approaches resonance, $\rho_{rr}$ increases and eventually one domain
(either A or B depending on $\Omega_{\textrm{pert}}$) reaches the
threshold $\rho_{\textrm{th}}$, this threshold effect is also simulated
by the epidemic model in the main text. At this detuning [label $g$
in Fig.~\ref{transmission}(a)], the system undergoes a non-equilibrium
phase transition accompanied by a sudden jump in its optical transmission.
Note that the critical detuning also depends on temperature and $\Omega_{p}$.
Increasing $\Delta_{c}$ further, the system initially follows the
steady-state 2 resonance curve, and then undergoes a second phase
transition at point $i$ in Fig.~\ref{transmission}(a) as $\rho_{rr}$
lies above threshold in both domains. In this example, the transition
points $g$ and $i$ do not coincide due to different local conditions
in the domains A and B. By changing $\Omega_{\textrm{pert}}$, $g$ and
$i$ can be moved to coincide. Scanning $\Delta_{c}$ from blue- to
red-detuning ($-$ scan direction, black line), the phase transitions
occur at points $d$ and $b$, giving rise to two hysteresis loops, see Fig.~\ref{transmission}(a). Depending on system parameters, the
two bistability loops may be either closed [Fig.~\ref{transmission}(a)]
or open. Three states with different Rydberg fractions, (for example $f_{R}<0.9$, $f_{R}=0.23$, and $f_{R}=0.33$, respectively), are used in the simulations. The system transmission states $T(f_{R})$
are treated as EIT transmission with different energy shifts and broadening,
given by the gray lines in Fig.~\ref{transmission}(a). The multiple
discrete transmission levels arise as the transition threshold is
reached for different detunings in each domain.

To highlight how the subtle difference in Rydberg population
between domains A, B, and C results in optical multistability, we analyze three
situations: no-, single- and double-bistability. The spatial variation of the perturbing, probe, and
coupling fields leads to a spatially inhomogeneous coupling to the
Rydberg state and creates an inhomogeneous Rydberg density. Once the critical
Rydberg population in a given region exceeds the threshold population $\rho_{\textrm{th}}$, the state of the region would transition
from the NI to the I phase, and accompanied by a jump in the probe transmission. If
the Rydberg population does not exceed the threshold population $\rho_{\textrm{th}}$
in any domain, we only observe a standard EIT spectrum without any jump,
as in the case in Fig.~\ref{transmission}(b). The probe transmission spectra with $+$ and $-$
scan directions are smooth without sudden change. If the Rydberg population in, e.g. region B, is above $\rho_{\textrm{th}}$, optical bistability occurs
as in the case in Fig.~\ref{transmission}(c). We only observe a single bistability
because there is only one domain. If the Rydberg population of both domains A and
B exceed $\rho_{\textrm{th}}$, but with different densities, we could observe optical
multistability in the spectra, in which the two bistabilities are staggered,
as shown in Fig.~\ref{transmission}(d).

\clearpage{}

\clearpage{}
\textbf{IV. Comparing experimental and theoretical data for the spatial domains}

\begin{figure}[b]
\includegraphics[width=0.88\columnwidth]{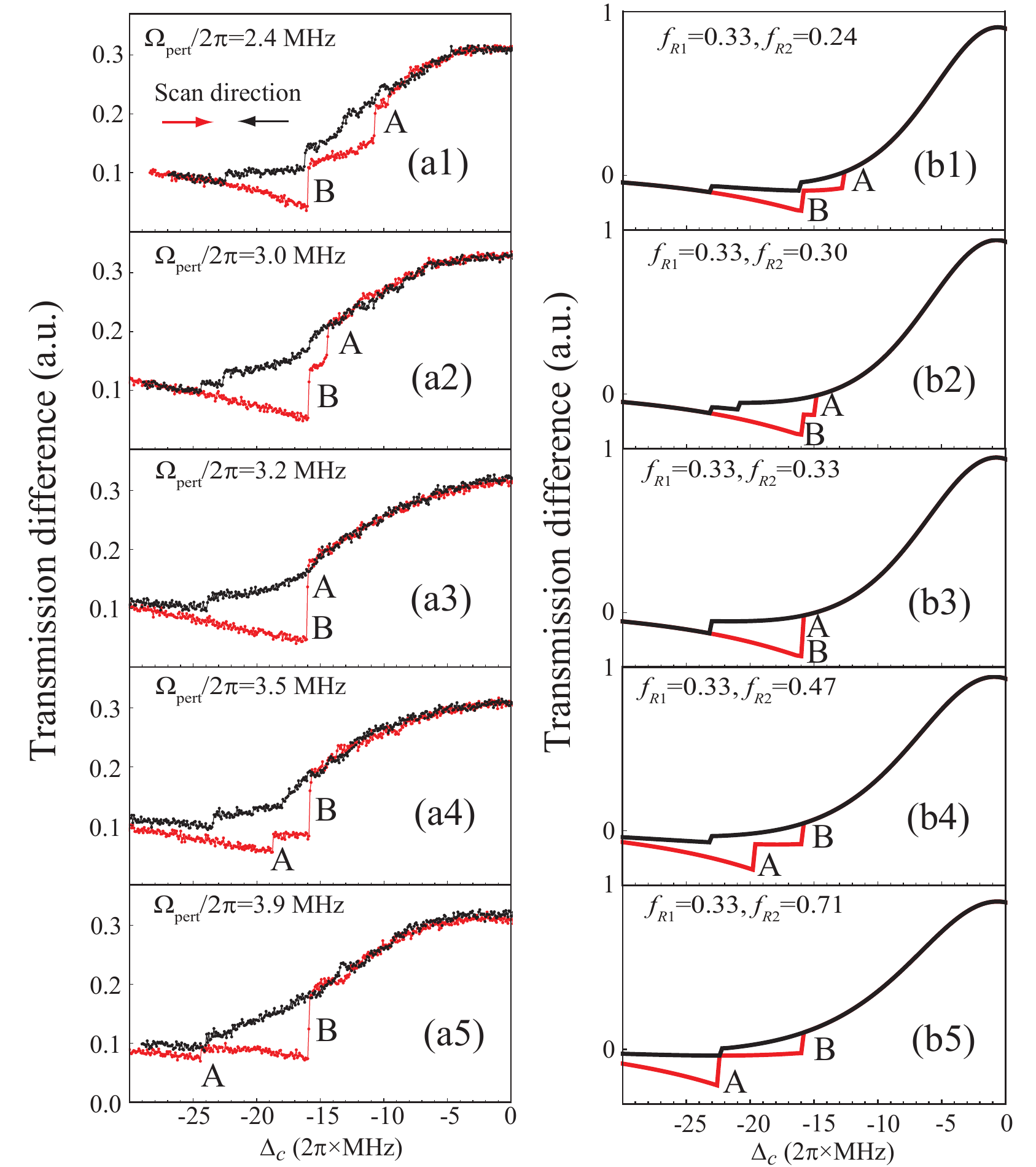}\caption{Effect of changing the perturbing field Rabi frequency $\Omega_{\textrm{pert}}$
on multi-domain bistability. Comparison of the theoretically (a1-a5) and
experimentally (b1-b5) observed probe transmission. Positive ($+$)
and negative ($-$) scan directions are shown in red and black, respectively.
The hysteresis loops for $\Omega_{\textrm{pert}}=(2\pi\times2.4,~3.0,~3.2,~3.5,~3.9)$
MHz are shown in Figs.~3(a1-a5), respectively. The phase transitions
associated with domains A and B for the $+$ scans are labelled accordingly.
Figures~b1-b5 show simulations based on our multi-domain model. In
the model, we use the Rydberg fractions $f_{R1}=0.33$ and $f_{R2}=0.24$
(b1), $f_{R2}=0.30$ (b2), $f_{R2}=0.33$ (b3), $f_{R2}=0.47$ (b4),
$f_{R2}=0.71$ (b5), where $f_{Rj}$ are the fractions of Rydberg
atoms in the interaction volumes A and B, respectively (see text). The
Rabi frequency of the probe is set to $\Omega_{p}=2\pi\times5.7$~MHz.
\label{multibistability-tuning}}
\end{figure}

In order to simulate the behavior of the spatial domains in the phenomenological
model, we assign distinct fractions $f_{R1}$ and $f_{R2}$ to
each domain and combine their overall optical response. In both theory
and experiments, we observe two bistable regimes, and a multistable
regime where the former coincide. For $\Omega_{\textrm{pert}}=2\pi\times2.4$
MHz [Fig.~\ref{multibistability-tuning}(a1)], we observe similar
behaviour predicted in Fig.~\ref{Experimental setup-1}(c) with thresholds
(+ scan directions) for state 2 at $\Delta_{c}\sim-2\pi\times16$
MHz (transition in domain B) and state 3 at $\Delta_{c}\sim-2\pi\times10$~MHz
(domain A). To demonstrate control of the phase transition point in
domain A, we vary $\Omega_{\textrm{pert}}$ from $2\pi\times2.4$
to $2\pi\times3.9$~MHz. Scanning $\Delta_{c}$ in either direction,
we obtain the experimental spectra shown in Figs.~\ref{multibistability-tuning}(a1-a5).
For $\Omega_{\textrm{pert}}=2\pi\times2.4$ MHz, Fig.~\ref{multibistability-tuning}(a1),
the hysteresis loops hardly overlap. For $\Omega_{\textrm{pert}}=2\pi\times3.2$~MHz,
Fig.~\ref{multibistability-tuning}(a3), the transition appears at
the same critical detuning as the phase transitions occur simultaneously
in both domains. Comparing Fig.~\ref{multibistability-tuning}(a1-a5)
and (b1-b5), the model and experimental data are in good qualitative
agreement. But if we choose $f_{R1}$ and $f_{R2}$ to reproduce the
experimental thresholds for the $+$ scan direction of $\Delta_{c}$, we
observe slight deviations between the theoretically predicted and experimentally
observed thresholds for the $-$ scan direction. If we scan $\Delta_{c}$
more slowly for the parameters in Fig.~\ref{multibistability-tuning}a3,
the double bistability appears again because of slight population
differences between domains due to fluctuations of the driving fields
or vapor temperature.

\clearpage{}

\clearpage{}

\textbf{V. Susceptible infected-susceptible and susceptible-infected-recovered  models}

\begin{figure}[h]
\includegraphics[width=1\columnwidth]{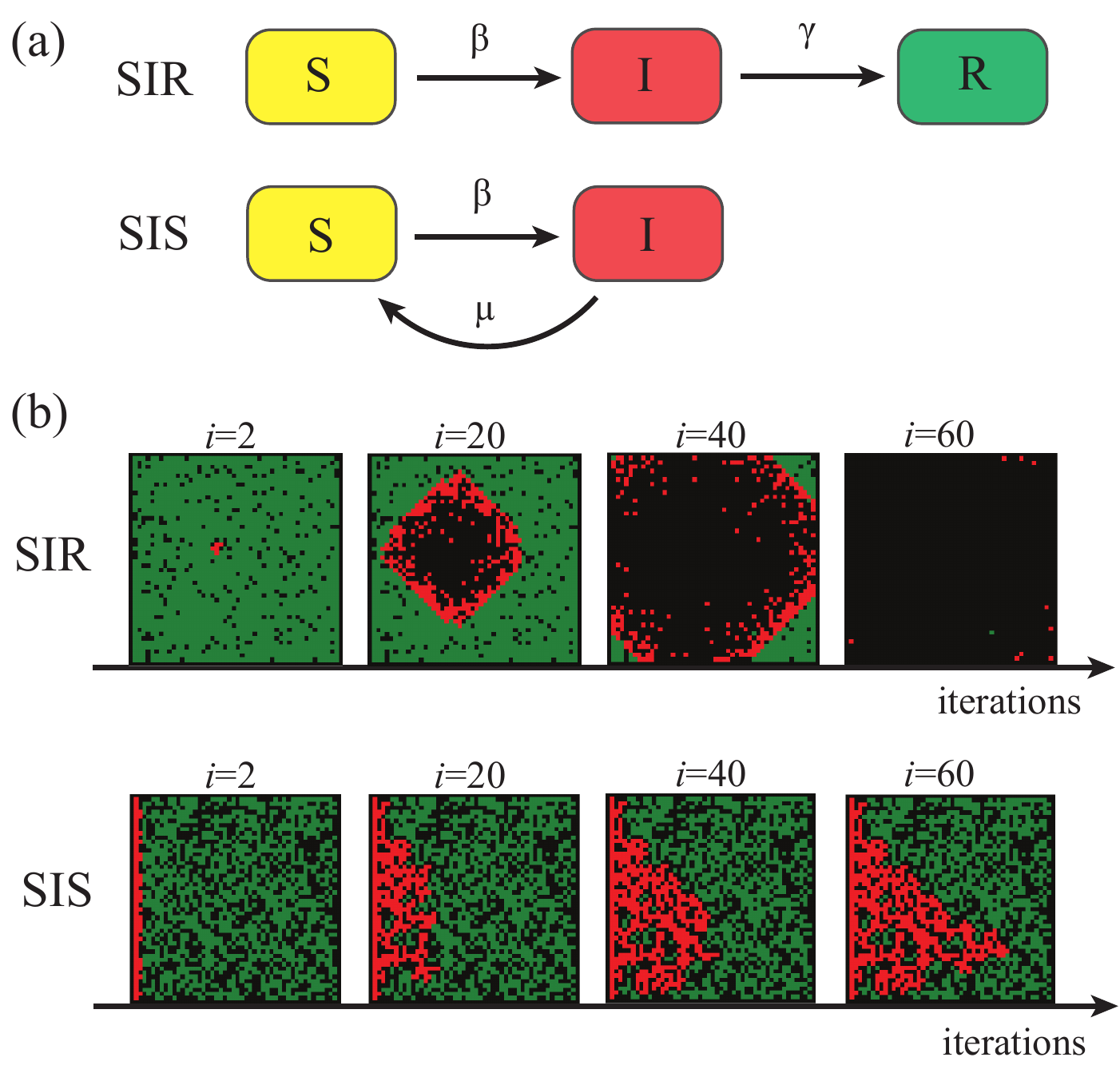}\caption{(a) Diagrammatic representation of the SIS and SIR models in terms of reaction-diffusion processes. Boxes represent different disease states, while the arrows represent the stochastical transitions between these states. Note that S: susceptible state; I: infected state; R: recovered state; $\beta$ represents the probability of the transition $S\rightarrow I$; $\gamma$ is the recovery probability from infected $I$ to recovered $R$ via isolation and medication; $\mu$ is the probability of the transition $I\rightarrow S$, corresponding to when the infected individuals have not been immunized. (b) A 2D snapshot of interacting Rydberg atoms spreading in a $50\times50$ system after iterations $i = 2, 20, 40, 60,$ respectively for $f_{R} = 0.9$ (SIR model) and $f_{R} = 0.55$ (SIS model). Red cells correspond to the infected state $I$, green cells the susceptible state $S$, and black cells the recovered state $R$.}

\label{SIS and SIR models}
\end{figure}

Two typical epidemic spreadings can be described by the susceptible-infected-susceptible (SIS) and susceptible-infected-recovered (SIR) models, as shown in Fig.~\ref{SIS and SIR models}(a). The SIS model has two simple transitions: (1) $S\rightarrow I$ happening when a susceptible individual S interacts with an infected I and becomes infected; (2) the reverse process $I\rightarrow S$ occurs when the infected individual I becomes susceptible. The SIS model does not have an immunity state, and remains in the permanent cycle $I\rightarrow S\rightarrow I$. The SIR model consists of three states: the susceptible S, infected I, and recovered states R. The main difference between the SIR and SIS modes is that $I\rightarrow S$ is replaced by $I\rightarrow R$ in SIR model. The process $I\rightarrow R$ corresponds to the infected individual I recovering to be immune R via isolation and medication.

In order to demonstrate the spatial evolution of the epidemic, we build a cellular automaton iteration algorithm as discussed in the main text. We investigate the dynamical results of the SIS and SIR models as we change the fractions $f_{R}$ of the susceptible state. We show the spreading dynamics of the infected state after different iterations $i=2,20,40,60,$ for different fractions $f_{R}$. For low fractions $f_{R}$, e.g. $f_{R}<0.2$, there is almost no spread because there are not enough susceptible individuals to match close neighbor contact, while for larger fractions $f_{R}$, i.e. $f_{R}=0.9$ shown in Fig.~\ref{SIS and SIR models}(b) (SIR model), the infected state is more inclined to spread through all the volume. We also observe the evolution of the SIS model with $f_{R}=0.55$: the infected individuals spread along local trajectories and result in a specific graph state [SIR model in Fig.~\ref{SIS and SIR models}(b)].

Eventually, the final states of evolution for these two models are different. For the SIS model, the system follows a stationary state due to the cycle $I\rightarrow S\rightarrow I$; see the evolution of the system after the iteration $i=60$ [SIS model in Fig.~\ref{SIS and SIR models}(b)]. This stationary effect satisfies the character of ``endemic state'' in which the dynamic equilibrium between the susceptible and infected state is sustained for a long time. While in the SIR model, due to the recovery channel $S\rightarrow R$, most of the red and green cells change to black cells after multiple iterations [SIR model in Fig.~\ref{SIS and SIR models}(b)], and the system finally reaches an immune state in which all infected individuals are healthy. This describes the process of ``herd immunity''.

\clearpage{}

\clearpage{}

\textbf{VI. Effects of fast and slow scan of the detuning}

\begin{figure}[h]
\includegraphics[width=0.75\columnwidth]{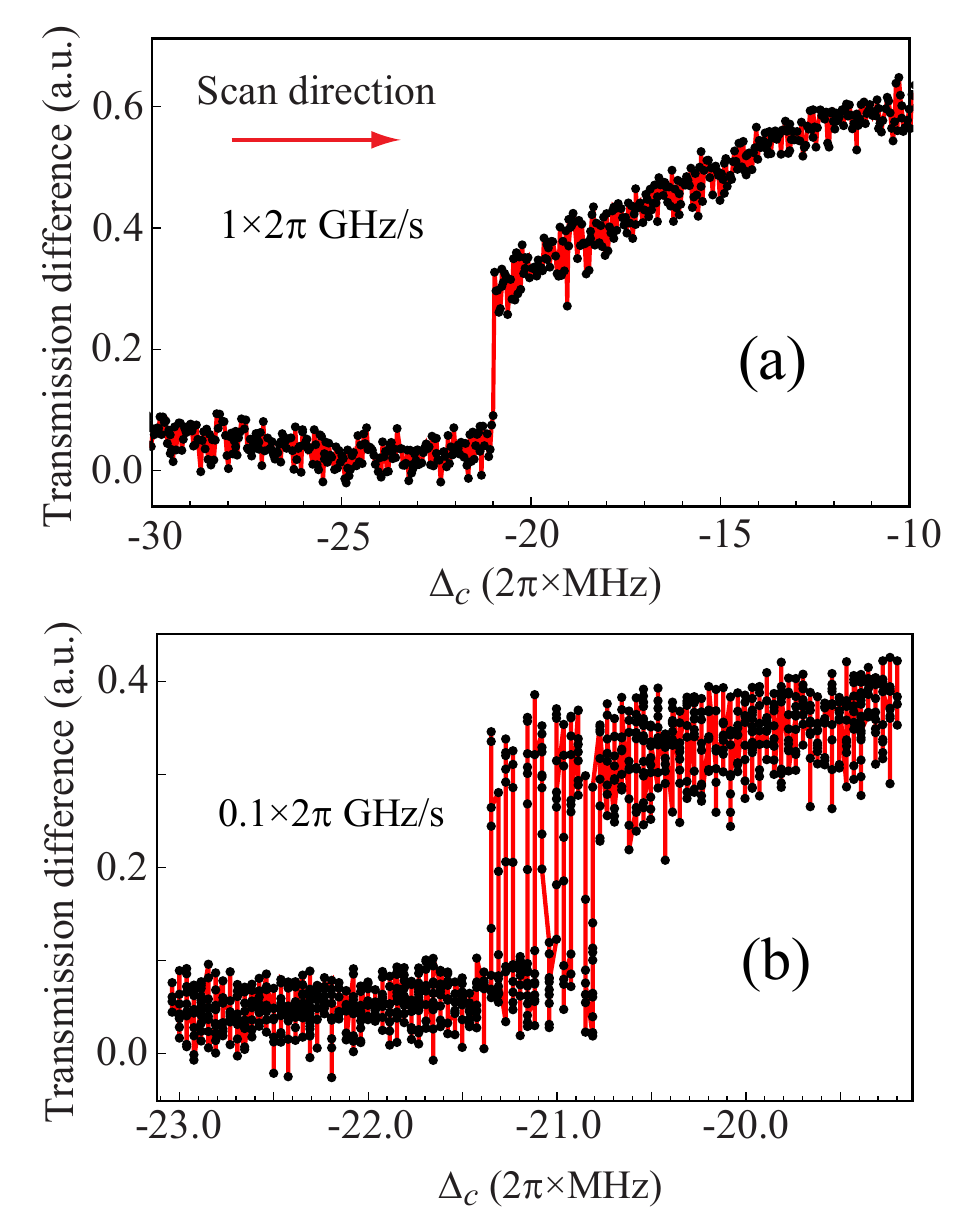}\caption{Transmission in the vicinity of the critical threshold for scan rates
of $1\times2\pi$~GHz/s (a) and $0.1\times2\pi$~GHz/s (b). For
slow scans, the Rydberg density can become insufficient to sustain
the interacting phase as the decay dominates over the change in coupling
strength to the Rydberg state from the scan.}

\label{scan rate2}
\end{figure}

At a low scan rate of $\Delta_{c}$, the spreading dynamics of the interacting phase
population is affected by the finite lifetime of the Rydberg state, as well as the interacting phase, compared to the sweep speed of the
coupling detuning $\Delta_{c}$ [Fig.~\ref{scan rate2}]. The interacting Rydberg
population would oscillate between $\Delta_{c}=2\pi\times-21.3$~MHz and $\Delta_{c}=2\pi\times-20.8$~MHz, which corresponds to the case when $\beta\Delta t\sim\gamma\Delta t$ and $\mu\Delta t\rightarrow0$ in the relatively longer time interval $\Delta t$, see Fig.~\ref{scan rate2}(b). Here, the increase in
coupling strength to the Rydberg state arising from the detuning scan
is low, and the interacting phase cannot be sustained because the
decay plays a role in the interacting phase spreading out. In addition, due to the decay of the interacting Rydberg atoms, and the refilling atoms from the thermal motions, then the system exhibits self-organized quasi-criticality \cite{klocke2020hydrodynamic},
making the system unstable in the interacting phase. Moreover,
a phase transition without oscillation appears at a fast scan rate of $\Delta_{c}$,
see Fig.~\ref{scan rate2}(a); in this case, the spreading rate $\beta$ is
larger than the decay rate $\gamma$, thus $\gamma\Delta t\rightarrow0$ or $\beta\Delta t\gg\gamma\Delta t,\mu\Delta t$ in a relatively short time interval $\Delta t$.

The system under the different regimes shows a different non-equilibrium dynamics. Our experiments in the fast-scan regime follow the predictions of the SIS model. In the fast-scan, the interacting Rydberg atoms do not have enough time to decay, because the increased Rydberg population would supply the system when scanning $\Delta_{c}$ fast. Although the interacting Rydberg atoms are constantly supplied, there are still some interacting atoms decaying into non-interacting atoms. But in a relatively long time, the system would reach a stationary state, where the interacting and the non-interacting Rydberg atoms are in dynamical equilibrium. This process follows the predictions of the SIS model: the disease does not confer immunity and individuals can be infected over and over again, undergoing a cycle of susceptible state $\rightarrow$ infected state $\rightarrow$  susceptible state, which, under some conditions, can be sustained in a long time \cite{pastor2015epidemic}. While in slow- or no-scan regime, the system follows the predictions of the SIR model. The system near the critical point would undergo a self-organized phase, where the lost atoms are not effectively replenished, thus the interacting phase would transit to the no-interacting phase. For relatively long times, the interacting Rydberg atoms
have enough time to be lost below the threshold $\rho_{\textrm{th}}$, and this corresponds to the process of infectious individuals recovering from the disease in the SIR model.

\hphantom{}

\clearpage{}

\bibliographystyle{ScienceAdvances}